\documentclass[12pt,a4paper]{article}
\usepackage{a4wide}
\usepackage{amsfonts}
\begin{document}
{\renewcommand{\thefootnote}{\fnsymbol{footnote}}
\hfill  PITHA -- 99/32\\
\medskip
\hfill gr--qc/9910103\\
\medskip
\begin{center}
{\LARGE  Loop Quantum Cosmology I: Kinematics }\\
\vspace{1.5em}
Martin Bojowald\footnote{e-mail address:
{\tt bojowald@physik.rwth-aachen.de}}\\
Institute for Theoretical Physics, RWTH Aachen\\
D--52056 Aachen, Germany\\
\vspace{1.5em}
\end{center}
}

\setcounter{footnote}{0}

\newtheorem{lemma}{Lemma}

\newcommand{\proofend}{\raisebox{1.3mm}{\fbox{\begin{minipage}[b][0cm][b]{0cm}
\end{minipage}}}}
\newenvironment{proof}{\noindent{\it Proof:} }{\mbox{}\hfill 
  \proofend\\\mbox{}}

\newcommand{\AbB}{\overline{{\cal A}}_B}
\newcommand{\UbB}{\overline{{\cal U}}_B}
\newcommand{\AbS}{\overline{{\cal A}}_{\Sigma}}
\newcommand{\Haux}{{\cal H}_{{\mathrm{aux}}}}

\newcommand{\md}{{\mathrm{d}}}
\newcommand{\Aut}{\mathop{\mathrm{Aut}}}
\newcommand{\Ad}{\mathop{\mathrm{Ad}}\nolimits}
\newcommand{\ad}{\mathop{\mathrm{ad}}\nolimits}
\newcommand{\Hom}{\mathop{\mathrm{Hom}}}
\newcommand{\Ima}{\mathop{\mathrm{Im}}}
\newcommand{\id}{\mathop{\mathrm{id}}}
\newcommand{\diag}{\mathop{\mathrm{diag}}}
\newcommand{\Kern}{\mathop{\mathrm{ker}}}
\newcommand{\tr}{\mathop{\mathrm{tr}}}
\newcommand{\sgn}{\mathop{\mathrm{sgn}}}
\newcommand{\semidir}{\mathrel{\mathrm{\times\mkern-3.3mu\protect%
\rule[0.04ex]{0.04em}{1.05ex}\mkern3.3mu\mbox{}}}}
\newcommand{\dive}{\mathop{\mathrm{div}}}
\newcommand{\Diff}{\mathop{\mathrm{Diff}}\nolimits}

\newcommand*{\R}{{\mathbb R}}
\newcommand*{\N}{{\mathbb N}}
\newcommand*{\Z}{{\mathbb Z}}
\newcommand*{\Q}{{\mathbb Q}}
\newcommand*{\C}{{\mathbb C}}

\begin{abstract}
  The framework of quantum symmetry reduction is applied to loop
  quantum gravity with respect to transitively acting symmetry groups.
  This allows to test loop quantum gravity in a large class of
  minisuperspaces and to investigate its features -- e.g.\ the
  discrete volume spectrum -- in certain cosmological regimes.
  Contrary to previous studies of quantum cosmology (minisuperspace
  quantizations) the symmetry reduction is carried out not at the
  classical level but on an auxiliary Hilbert space of the quantum
  theory before solving the constraints. Therefore, kinematical
  properties like volume quantization survive the symmetry reduction.
  In this first part the kinematical framework, i.e.\ implementation
  of the quantum symmetry reduction and quantization of Gau\ss\ and
  diffeomorphism constraints, is presented for Bianchi class A models
  as well as locally rotationally symmetric and spatially isotropic
  closed and flat models.
\end{abstract}

\section{Introduction}

One of the main applications of general relativity has always been the
study of cosmological models, i.e.\ of solutions which allow a
transitive group of space isometries. This symmetry condition reduces
the infinite number of degrees of freedom of general relativity to
finitely many ones for these homogeneous minisuperspace models
\cite{DeWitt,Misner}, leading to an extensive use as test models for a
quantum theory of gravity. In this respect they are similar to quantum
mechanical models with the role of the Schr\"odinger equation played
by the Wheeler--DeWitt equation \cite{DeWitt} which is the quantized
Hamiltonian constraint of general relativity. This equation is a
hyperbolic differential equation involving the scale factor of the
universe, which plays a role analogously to a time variable. It
remains hyperbolic after small perturbations of the homogeneous
metrics \cite{Giulini}.

In lack of a complete quantum theory of gravity this approach of
quantization {\em after\/} symmetry reduction has long been the only
possibility to study quantum effects in cosmology, which are believed
to have a large impact on the development of a universe, at least in
very early stages. But now there are candidates for such a
quantum theory with powerful techniques, and it is legitimate to ask
what these theories have to say about such questions. The most ambituous
approach which is claimed to provide a quantum theory of gravity is
string theory \cite{Polchinski}. In the cosmological context it has
been applied to the study of inflationary models because of its field
content different from general relativity \cite{Effective,
  Background,StringCos}. A second novelty is a scale factor duality
\cite{Veneziano} which relates universes of small and large scale. But
the approach to quantum cosmology is the same as that of general
relativity, only the Einstein-Hilbert action is changed by some
effective terms. Thus, concerning quantization of the metric up to now
nothing conceptually new comes into play by using ideas of string
theory.

In a second approach to quantum gravity, loop quantum gravity
\cite{Rov:Loops}, the situation is different. Basic geometric
quantities have been quantized \cite{AreaVol,Loll:Vol,Area,Vol2,Len}
and found to have discrete spectra. Their eigenstates are spin network
states, a discrete structure of space. Continuous space is regarded as
an approximate concept at large scales, which can be described by
weave states \cite{Weaves}. Some preliminary considerations in the
cosmological context using these weaves have appeared in reference
\cite{Smolin}. Evidently, in such a situation concepts like a
hyperbolic differential equation with respect to the scale factor
cannot remain true. The departure from those ideas of differential
Wheeler--DeWitt equations can best be seen by looking at the
quantization of the Wheeler--DeWitt operator in loop quantum gravity
\cite{AnoFree,QSDI,QSDII}. Unfortunately, this operator of the full
theory remains poorly understood.

In the present paper we make use of symmetric, distributional states
of loop quantum gravity which have been defined and investigated in
reference \cite{SymmRed}. Contrary to weaves, they are exactly
symmetric, not only approximately at large scales. This fact
guarantees that the number of degrees of freedom is reduced to
finitely many ones as in classical symmetry reductions. Nevertheless,
we impose the symmetry conditions in quantum theory, namely in the
auxiliary Hilbert space of loop quantum gravity, and we can use all
the techniques of loop quantum gravity for the investigation of
cosmological models. In particular, spectra of geometric operators
remain discrete. In these reduced models we then have to quantize and
solve the reduced constraints, which will be done in the present part
for the kinematical ones. Concerning the more complicated Hamiltonian
constraint, which leads to the Wheeler--DeWitt equation, our reduced
models can provide helpful tests for its quantization
\cite{cosmoIII} as well as solution.

In this first part the emphasis lies on the implementation of the
kinematical framework. Besides providing the stage for future work
this will serve us as a means to test some ideas of the symmetry
reduction of reference \cite{SymmRed}. In particular, the Higgs
constraint (\cite{SymmRed} and Section~\ref{s:symmred}) can be
analysed more easily because there is just one Higgs vertex in the
reduced model. Cosmological considerations will not appear in this
part, and therefore we will not bother ourselves with matter
couplings.

The next section recalls the necessary material of reference
\cite{SymmRed} specialized to transitive group actions.
Section~\ref{s:models} introduces some cosmological models classically
as well as in its quantum symmetry reduced form. In
Section~\ref{s:gauss} we quantize and solve the Gau\ss\ constraints
and in Section~\ref{s:diff} the diffeomorphism constraints for these
models.

\section{Quantum Symmetry Reduction for Transitive\\ Symmetry Groups}
\label{s:symmred}

In this section we provide the mathematical prerequisites for a quantum
treatment of cosmological models. These are the classification of
symmetric principal fiber bundles and invariant connections thereon
\cite{KobNom} and the general framework of quantum symmetry reduction
\cite{SymmRed}, both specialized to transitive actions of a symmetry
group. This section thereby serves to fix our notation.

\subsection{Invariant Connections}

Let $P(\Sigma,G,\pi)$ be a principal fiber bundle over the compact
manifold $\Sigma$, which is regarded as a space manifold for canonical
quantization, and with structure group $G$, which will be $G=SU(2)$
for gravity formulated in real Ashtekar variables. On $P$ there is a
given transitively acting symmetry group $S<\Aut P$ of bundle
automorphisms.  To allow for group actions with rotational symmetry in
addition to homogeneity, $S$ can have a non-trivial isotropy (this
mathematical notion of isotropy should not be confused with the
physical concept) subgroup $F<S$ (fixing a point $x_0$ in $\Sigma$),
which is up to conjugacy the same for all points in $\Sigma$ due to
transitivity of the action of $S$. $\Sigma$ can be represented as
$\Sigma\cong S/F$, and $\Sigma/S=:B=\{x_0\}$ is represented by a
single point which can be chosen arbitrarily in $\Sigma$. The general
framework demands that the coset space $S/F$ is reductive
\cite{KobNom}, i.e.\ the Lie algebra of $S$ can be decomposed as
${\cal L}S={\cal L}F\oplus{\cal L}F_{\perp}$ with $\Ad_F{\cal
  L}F_{\perp}\subset{\cal L}F_{\perp}$. Important examples are
semisimple groups $S$, in which case ${\cal L}F_{\perp}$ is the
orthogonal complement of ${\cal L}F$ with respect to the
Cartan-Killing metric, freely acting groups with $F=\{1\}$ and ${\cal
  L}F_{\perp}={\cal L}S$, and semidirect products $S=N\semidir F$ with
${\cal L}F_{\perp}={\cal L}N$. We will encounter the last case when
studying homogeneous models with a rotational isotropy subgroup $F$.
$N$ will then be a translational subgroup (isomorphic to one of the
Bianchi groups) of $S$, and act freely and
transitively.

The isotropy subgroup plays an important role in the classification of
symmetric bundles and invariant connections \cite{KobNom}. It provides
in the first place a map $F\colon\pi^{-1}(x_0)\to\pi^{-1}(x_0)$, by
means of which group homomorphisms $\lambda_p\colon F\to G$ can be
defined, for each point $p$ in the fiber over $x_0$, by means of
$f(p)=:p\cdot\lambda_p(f)$ for all $f\in F$. By choosing a different
point $p\cdot g$ in the fiber over $x_0$ such a homomorphism gets
conjugated: $\lambda_{p\cdot g}=\Ad_{g^{-1}}\circ\lambda_p$.
Therefore, for the following classifications only the conjugacy class
$[\lambda]$ of a given homomorphism matters. But for homomorphisms
from different conjugacy classes the $S$-actions on a given principal
fiber bundle $P$ are inequivalent, and therefore all $S$-symmetric
principal fiber bundles are classified by a conjugacy class
$[\lambda]$ of group homomorphisms $\lambda\colon F\to G$.

The next question is, given an $S$-symmetric principal fiber bundle $P$
classified by $[\lambda]$, what is the general form of a
$[\lambda]$-invariant, i.e.\ invariant with respect to this classified
action of $S$ on $P$, connection on $P$. By using the Maurer--Cartan
form $\theta_{\mathrm{MC}}$ on $S$ and an embedding $\iota\colon
S/F\hookrightarrow S$ all such connections can be written in the form
\begin{equation}
  \omega_{S/F}=\phi\circ\iota^*\theta_{\mathrm{MC}} \label{recons}
\end{equation}
where $\phi\colon{\cal
  L}F_{\perp}\to{\cal L}G$ is a linear map obeying the equation
\begin{equation}\label{Higgs}
  \phi(\Ad_f(X))=\Ad_{\lambda(f)}\phi(X)
\end{equation}
for all $f\in F$, $X\in{\cal L}F_{\perp}$, and where
$\lambda\in[\lambda]$ is chosen from the conjugacy class. In what
follows the map $\phi$ will be denoted as Higgs field. The structure
group $G$ acts on $\phi$ by conjugation, which stems from the usual
gauge transformation of a connection; the solution space of equation 
(\ref{Higgs}) is, however, invariant only with respect to the reduced
structure group $Z_{\lambda}:=Z_G(\lambda(F))$, the centralizer of
$\lambda(F)$ in $G$. This fact leads to a partial gauge fixing which
is manifest in all classical symmetry reductions: The reconstructed
connection form $\omega_{S/F}$ is a $Z_{\lambda}$-connection and will
in general depend explicitly on $\lambda$. As noted above, only the
conjugacy class $[\lambda]$ plays a gauge invariant role, and indeed
after choosing a different $\lambda'\in[\lambda]$ we would reconstruct
a gauge equivalent connection. In classical symmetry reductions a
fixed $\lambda\in[\lambda]$ is chosen once and for all leading to a
breaking of the structure group from $G$ to $Z_{\lambda}$.

\subsection{Symmetric States}

The basic idea of reference \cite{SymmRed} is to use the
reconstruction of invariant connections from the Higgs field to pull
back a spin network function, which is a function on the space $\AbS$
of generalized connections on $\Sigma$ modulo gauge transformations,
to a function on the space $\AbB\times\UbB$ of generalized connections
and Higgs fields on the reduced manifold $B$.  It was also shown
there, how the functions on $\AbB\times\UbB$ can be interpreted as
distributional states of the unreduced theory which are
$[\lambda]$-symmetric in the sense that their supports contain only
$[\lambda]$-invariant connections on $\Sigma$. Furthermore, it was
shown that all symmetric states can be obtained in such a way.

In the case of transitively acting symmetry groups, $B=\{x_0\}$
consists of a single point and the space $\UbB$ is finite-dimensional,
whereas $\AbB$ certainly makes no contribution. The reduced
theory will hence only have finitely many degrees of freedom after a
quantum symmetry reduction (analogously to the classical reduction).

There are some subtleties because of the partial gauge fixing which
will show up in the solution space of equation (\ref{Higgs}). To start
with, for each basis element of ${\cal L}F_{\perp}$ we will use a
separate Higgs field component which will be described by using point
holonomies \cite{FermionHiggs} in the quantum theory. By using point
holonomies with respect to the structure group $G$ instead of
$Z_{\lambda}$ the partial gauge fixing can be undone. This is even
mandatory, because point holonomies necessarily transform under the
adjoint representation of the structure group, whereas Higgs fields
will not necessarily transform under the adjoint representation of
$Z_{\lambda}$. The space of point holonomies will, however, be the
quantum configuration space only for freely acting symmetry groups. If
$F$ is non-trivial, the condition (\ref{Higgs}) will further constrain
this space. More details will be given in the next section, and
concerning solutions of equation (\ref{Higgs}) in the next part of
this series \cite{cosmoII}.

\section{Bianchi Class A Models, Locally Rotationally\\ Symmetric and
  Isotropic Models}
\label{s:models}

Here we introduce the models we are interested in: Bianchi class A
models constitute all homogeneous models with a freely acting symmetry
group ($F=\{1\}$) which can be treated in a Hamiltonian formulation
(Bianchi class B models violate the principle of symmetric criticality
\cite{classAB,midisup}). Some of them can be reduced further on by
demanding rotational symmetry with one axis ($F=U(1)$) or even
isotropy ($F=SU(2)$, in general this isotropy subgroup does not
project to an $SO(3)$-action on $P$, although it does on $\Sigma$).

\subsection{Bianchi Class A Models}

Bianchi models describe all possible types of metrics which are
homogeneous in space \cite{MacCallum}. They have been discussed in a
minisuperspace quantization e.g.\ in reference \cite{MiniQuant}. The
classical reduction in terms of complex Ashtekar variables has been
carried out in reference \cite{Kodama}. Here we present the reduction
for real Ashtekar variables in the framework described in the
preceding section.

In the context of Bianchi models the transitive symmetry group acts
freely on $\Sigma$, which implies that $\Sigma$ can be identified with
the group manifold $S$ (up to a suitable compactification if $S$ is
non-compact). The three generators of ${\cal L}S$ are denoted as $T_I$,
$1\leq I\leq 3$, with the relations $[T_I,T_J]=c^K_{IJ}T_K$. Here
$c^K_{IJ}$ are the structure constants of ${\cal L}S$ fulfilling
$c^J_{IJ}=0$ for class A models by definition. The Maurer--Cartan form
on $S$ is given by $\theta_{\mathrm{MC}}=\omega^IT_I$ with left invariant
one-forms $\omega^I$ on $S$ which fulfill the Maurer--Cartan equations
\begin{equation}\label{MC}
  \md\omega^I=-{\textstyle\frac{1}{2}}c^I_{JK}\omega^J\wedge\omega^K\,.
\end{equation}
Due to $F=\{1\}$ all homomorphisms $\lambda\colon F\to G$ are given by
$1\mapsto 1$, and we can use the embedding $\iota=\id\colon
S/F\hookrightarrow S$. An invariant connection then takes the form
$A=\phi\circ\theta_{\mathrm{MC}}=\phi_I^i\tau_i\omega^I=A_a^i\tau_i\md
x^a$ with the matrices $\tau_j=-\frac{i}{2}\sigma_j$ generating ${\cal
  L}SU(2)$ ($\sigma_j$ are the Pauli matrices). The Higgs field is
given by $\phi\colon{\cal L}S\to{\cal
  L}G,T_I\mapsto\phi(T_I)=:\phi^i_I\tau_i$ already in its final form,
because the Higgs condition (\ref{Higgs}) is empty. By using the left
invariant vector fields $X_I$ obeying $\omega^I(X_J)=\delta^I_J$ and
with Lie brackets $[X_I,X_J]=c^K_{IJ}X_K$ the momenta canonically
conjugate to $A_a^i=\phi^i_I\omega^I_a$ can be written as
$E^a_i=\sqrt{g_0}\,p^I_iX^a_I$ with $p^I_i$ being canonically
conjugate to $\phi^i_I$. Here $g_0=\det(\omega^I_a)^2$ is the
determinant of the left invariant metric
$(g_0)_{ab}:=\sum_I\omega^I_a\omega^I_b$ on $\Sigma$ which is used to
provide the density weight of $E^a_i$. The symplectic
structure can be derived from
\[
  (\kappa\iota)^{-1}\int_{\Sigma}\md^3x\,\dot{A}^i_aE^a_i=
  (\kappa\iota)^{-1}\int_{\Sigma}\md^3x\,\sqrt{g_0}\,
  \dot{\phi}^i_I p^J_i \omega^I(X_J)=
  \frac{V_0}{\kappa\iota}\dot{\phi}^i_Ip^I_i\, ,
\]
to obtain
\begin{equation}
  \{\phi^i_I,p^J_j\}=\kappa\iota'\delta^i_j\delta^J_I
\end{equation}
with the gravitational constant $\kappa$ and the modified Immirzi
parameter $\iota':=\iota V_0^{-1}$ in which we absorbed the volume
$V_0:=\int_{\Sigma}\md^3x\sqrt{g_0}$ of $\Sigma$ measured in the
invariant metric $g_0$.

We proceed now by inserting the invariant connections and canonical
dreibeine into the Gau\ss, vector and Hamiltonian constraints. But
before doing so we show that the divergence of the density weighted
vector fields $\sqrt{g_0}X_I$, which appear in the Gau\ss\ constraint,
vanishes. To that end we use a metric independent duality
transformation which assigns to an $n$-form
$\omega=(n!)^{-1}\omega_{a_1\ldots a_n}\md
x^{a_1}\wedge\cdots\wedge\md x^{a_n}$ in $D$ dimensions the components
\[
  (*\omega)^{a_{n+1}\ldots a_D}:= (n!)^{-1}
    \epsilon^{a_1\ldots a_n a_{n+1}\ldots a_D} \omega_{a_1\ldots a_n}
\]
of a density weighted antisymmetric tensor, and, vice versa, to a
density weighted antisymmetric tensor $X$ the
differential form
\[
  *X:=[(D-n)!]^{-1} \epsilon_{a_1\ldots a_n a_{n+1}\ldots a_D}
  X^{a_1\ldots a_n} \md x^{a_{n+1}}\wedge\cdots\wedge\md x^{a_D}\,.
\]
Here, $\epsilon^{a_1\ldots a_D}$ and $\epsilon_{a_1\ldots a_D}$ are
the metric independent $\epsilon$-tensors in $D$ dimensions with
density weight $1$ and $-1$, respectively. The divergence of
$\sqrt{g_0}X_I$ can now be written as
$\dive\sqrt{g_0}X_I=*\md*(\sqrt{g_0}X_I)=*\md B_I$ with
\begin{eqnarray*}
  B_I & := & *(\sqrt{g_0}X_I)={\textstyle\frac{1}{2}}
  \epsilon_{abc}\sqrt{g_0}X_I^a\md 
    x^b\wedge\md x^c\\
  & = & {\textstyle\frac{1}{2}} \epsilon_{JKL}\omega_a^J\omega_b^K 
    \omega_c^L X_I^a\md x^b\wedge\md x^c = 
    {\textstyle\frac{1}{2}} \epsilon_{IKL}\omega^K_b \omega^L_c\md
    x^b\wedge\md x^c\\
  & = &{\textstyle\frac{1}{2}}\epsilon_{IKL}\omega^K\wedge\omega^L\,.
\end{eqnarray*}
The exterior derivative of $B_I$ can be calculated by using the
Maurer--Cartan equations (\ref{MC}):
\begin{eqnarray*}
  \md B_I & = & -\epsilon_{IKL}\omega^K\wedge\md\omega^L=
    {\textstyle\frac{1}{2}}\epsilon_{IKL}
    c^L_{MN}\omega^K\wedge\omega^M\wedge\omega^N\\
  & = & {\textstyle\frac{1}{2}}c^L_{MN}\epsilon_{IKL}\epsilon^{KMN}
  \sqrt{g_0}\,\md^3x= {\textstyle\frac{1}{2}}c^L_{MN}
  (\delta^M_L\delta^N_I -
  \delta^M_I\delta^N_L) \sqrt{g_0}\,\md^3x=-c^L_{IL}\sqrt{g_0}\,\md^3x\,.
\end{eqnarray*}
In the last two calculations we used the identity
\[
  \sqrt{g_0}={\textstyle\frac{1}{6}}
  \epsilon_{IJK}\epsilon^{abc}\omega^I_a\omega^J_b\omega^K_c 
\]
for the determinant of the invariant metric in terms of left invariant
one-forms. The divergence of $\sqrt{g_0}X_I$ can now be read off as
$\dive\sqrt{g_0}X_I=*\md B_I=-c^L_{IL}\sqrt{g_0}$, i.e.\ it vanishes
precisely for Bianchi class A models.

The Gau\ss\ constraint can now be calculated easily:
\begin{eqnarray}\label{gaussbianchi}
  {\cal G}_i & = & (\kappa\iota)^{-1}\int_{\Sigma}\md^3x
  (\partial_aE^a_i+\epsilon_{ijk}A^j_aE^a_k)=
  (\kappa\iota)^{-1}\int_{\Sigma}\md^3x (p^I_i\dive\sqrt{g_0}X_I+
  \sqrt{g_0}\epsilon_{ijk}\phi^j_Ip^I_k)\nonumber\\
  & = & (\kappa\iota')^{-1}(\epsilon_{ijk} \phi^j_Ip^I_k-c^J_{IJ})\,.
\end{eqnarray}

For the next two constraints we will need the curvature of
$A=\phi^i_I\tau_i\omega^I$. It is given by
\begin{eqnarray*}
  F & = & \md A+{\textstyle\frac{1}{2}}[A,A]= \phi^i_I\tau_i\md\omega^I+
   {\textstyle\frac{1}{2}}\epsilon_{ijk}
   \phi^i_I\phi^j_J\tau^k\omega^I\wedge\omega^J\\
  & = & {\textstyle\frac{1}{2}}
  (-\phi^i_Ic^I_{JK}+\epsilon_{ijk}\phi^j_J\phi^k_K)
  \tau_i \omega^J\wedge\omega^K
\end{eqnarray*}
using again the Maurer--Cartan equations. The components of the
curvature are
\[
  F^i_{IJ}=-\phi^i_Kc^K_{IJ}+\epsilon_{ijk}\phi^j_I\phi^k_J\,.
\]
They are now used to calculate the vector constraint with a shift
vector $N^a=N^IX_I^a$, $N^I$ constant (the fact that the $N^I$ are
constant on $\Sigma$ is a manifestation of a partial gauge fixing of
diffeomorphisms by demanding them to respect the symmetry; this
corresponds to choosing a special system of coordinates adapted to the
symmetry):
\begin{eqnarray}
 {\cal V}_aN^a & = & (\kappa\iota)^{-1}\int_{\Sigma}\md^3x\,
 F^i_{IJ}E^J_iN^I= (\kappa\iota)^{-1}\int_{\Sigma}\md^3x
 \sqrt{g_0}(-c^K_{IJ}\phi^i_Kp^J_i+
 \epsilon_{ijk}\phi^j_I\phi^k_Jp^J_i)N^I\nonumber\\
 & = & (\kappa\iota')^{-1}(-c^K_{IJ}\phi^i_Kp^J_i+\phi^j_I{\cal G}_j)N^I
\end{eqnarray}
where the first term
\begin{equation}\label{diffbianchi}
  {\cal D}_aN^a=-(\kappa\iota')^{-1}c^K_{IJ}\phi^i_Kp^J_iN^I
\end{equation}
is the diffeomorphism constraint. We note here that ${\cal G}$ and
${\cal D}$ are very similar: ${\cal G}$ generates as gauge
transformations conjugation in the internal $SU(2)$ space, whereas
${\cal D}$ generates conjugation in the homogeneous space $S$.
Therefore, the constants appearing in the constraints are the
structure constants $\epsilon_{ijk}$ of $SU(2)$ and $c^I_{JK}$ of $S$,
respectively. We will say more about this point when we quantize and
solve the diffeomorphism constraint in Section~\ref{s:diff}.

Finally, we calculate the Euclidean part of the Hamiltonian
constraint. In real Ashtekar variables there is an additional term in
the Lorentzian constraint \cite{AshVarReell}, which can, however, be
dealt with by the same methods as in reference \cite{QSDI} for the
full theory.  Therefore, we will not need its reduction explicitly.
The Euclidean part (with density weight 2) is given by
\begin{eqnarray}
  {\cal H}^{(E)} & = & \epsilon_{ijk}F^i_{IJ}E^I_jE^J_k=
  g_0(-\epsilon_{ijk} c^K_{IJ}\phi^i_Kp^I_jp^J_k+
  \epsilon_{ijk}\epsilon_{ilm} \phi^l_I\phi^m_Jp^I_jp^J_k)\nonumber\\
  & = & g_0(-\epsilon_{ijk}c^K_{IJ} \phi^i_Kp^I_jp^J_k+
  \phi^j_I\phi^k_Jp^I_jp^J_k-
  \phi^k_I\phi^j_Jp^I_jp^J_k)\,.\label{hambianchi}
\end{eqnarray}

\subsection{Locally Rotationally Symmetric and Isotropic Models}

On some of the Bianchi models additional symmetry conditions can be
imposed. If there is an isotropy subgroup $F\cong U(1)$ of the
symmetry group $S$, one obtains locally rotationally symmetric models
(LRS models, \cite{Ellis,MacCallum}; we use the term LRS only for the
restricted class of $F=U(1)$-models). Bianchi type I and IX can even
be constrained to isotropic metrics, i.e.\ $F\cong SU(2)$. (The only
other class A model for which this is possible is type
${\mathrm{VII}}_0$. It yields, however, an isotropic model equivalent
to that of type I \cite{MacCallum}.) These two models will be most
interesting in the following, because we can successively increase the
symmetry and observe properties of the quantum symmetry reduction step
by step. The essential idea of the quantum symmetry reduction of
reference \cite{SymmRed} is to pull back a function on the
unconstrained space of connections to a function on the space of
invariant connections by means of the reconstruction map
(\ref{recons}).  For the models considered here the pull back can be
decomposed into a map which leads to a function on the space of
homogeneous, but in general anisotropic, connections (this is the
quantum symmetry reduction for Bianchi models) followed by a map which
restricts the support of a function on the space of homogeneous
connections to only rotationally invariant ones. The second map, which
can be viewed as a symmetry reduction of its own, provides us with a
good test place for some of the ideas of quantum symmetry reduction.

These models with enhanced symmetry can be treated on an equal footing
by writing the symmetry group as the semidirect product
$S=N\semidir_{\rho}F$, with the isotropy subgroup $F$ and the
translational subgroup $N$, which is one of the Bianchi groups. The
composition in this group is defined as
$(n_1,f_1)(n_2,f_2):=(n_1\rho(f_1)(n_2),f_1f_2)$ which depends on the
group homomorphism $\rho\colon F\to\Aut N$ into the automorphism group
of $N$ (which for ease of notation will be denoted by the same letter
as the representation on $\Aut {\cal L}N$ used below). Inverse
elements are given by $(n,f)^{-1}=(\rho(f^{-1})(n^{-1}),f^{-1})$. To
determine the form of invariant connections we have to calculate the
Maurer--Cartan form on $S$ (using the usual notation):
\begin{eqnarray*}
  \theta^{(S)}_{\mathrm{MC}}(n,f) & = & (n,f)^{-1}\md
  (n,f)=(\rho(f^{-1})(n^{-1}),f^{-1})(\md n,\md f)\\
  & = & (\rho(f^{-1})(n^{-1})\rho(f^{-1})(\md n),f^{-1}\md f)=
  (\rho(f^{-1})(n^{-1}\md n),f^{-1}\md f)\\
  & = & (\rho(f^{-1})(\theta^{(N)}_{\mathrm{MC}}(n)),
    \theta^{(F)}_{\mathrm{MC}}(f))\,.
\end{eqnarray*}
Here the Maurer--Cartan forms $\theta^{(N)}_{\mathrm{MC}}$ on $N$ and
$\theta^{(F)}_{\mathrm{MC}}$ on $F$ appear. We next have to choose an
embedding $\iota\colon S/F=N\hookrightarrow S$, which can most easily
be done as $\iota\colon n\mapsto (n,1)$. We then have
$\iota^*\theta^{(S)}_{\mathrm{MC}}=\theta^{(N)}_{\mathrm{MC}}$, and a
reconstructed connection takes the form
$\phi\circ\iota^*\theta^{(S)}_{\mathrm{MC}}=\phi^i_I\omega^I\tau_i$
which is the same as for anisotropic models of the last subsection
(where now $\omega^I$ are left invariant one-forms on the translation
group $N$).  However, here $\phi$ is constrained by equation
(\ref{Higgs}) and we get only a subset as isotropic connections.

To solve the Higgs equation we have to treat LRS and isotropic models
separately. In the first case we choose ${\cal
  L}F=\langle\tau_3\rangle$, whereas in the second case we have ${\cal
  L}F=\langle\tau_1,\tau_2,\tau_3\rangle$ ($\langle\cdot\rangle$
denotes the linear span). Equation (\ref{Higgs}) can be written
infinitesimally as
\[
 \phi(\ad_{\tau_i}(T_I))=\ad_{\md\lambda(\tau_i)}\phi(T_I)=
 [\md\lambda(\tau_i),\phi(T_I)]
\]
($i=3$ for LRS, $1\leq i\leq 3$ for isotropy).  The $T_I$ denote the
generators of ${\cal L}N={\cal L}F_{\perp}$, on which the isotropy
subgroup $F$ acts as rotation: $\ad_{\tau_i}(T_I)=\epsilon_{iIK}T_K$.
This is the derivative of the representation $\rho$ defining the
semidirect product $S$. The conjugation on the left hand side of the
Higgs equation (\ref{Higgs}) is
$\Ad_{(1,f)}(n,1)=(1,f)(n,1)(1,f^{-1})=(\rho(f)(n),1)$, which follows
from the composition in $S$.

We next have to determine the possible conjugacy classes of
homomorphisms $\lambda\colon F\to G$. For LRS models their
representatives are given by $\lambda_k\colon U(1)\to SU(2),\exp
t\tau_3\mapsto\exp kt\tau_3$ for $k\in\N_0=\{0,1,\ldots\}$ (for a
derivation see the example of spherical symmetry in reference
\cite{SymmRed} which is in many respects similar to LRS models).
Choosing these representatives for $[\lambda_k]$ will be called
$\tau_3$-gauge. For the components $\phi^i_I$ of $\phi$ defined by
$\phi(T_I)=\phi^i_I\tau_i$ the Higgs equation takes the form
$\epsilon_{3IK}\phi^j_K=k\epsilon_{3lj}\phi^l_I$. This can be written
as a matrix equation $E_3\Phi=k\Phi E_3$ with
$(E_3)_{ij}:=\epsilon_{3ij}$ and $(\Phi)_{ij}:=\phi^j_i$ (indeed the
Higgs equation can be interpreted as an equation for $\phi$ to be an
intertwiner between certain subrepresentations of the representation
of $F$ on ${\cal L}F_{\perp}$ and the adjoint representation of $G$
\cite{Kubyshin}). This equation has a non-trivial solution only for
$k=1$, in which case $\phi$ can be written as
\[
  \phi_1=2^{-\frac{1}{2}}(a\tau_1+b\tau_2)\quad,\quad
  \phi_2=2^{-\frac{1}{2}}(-b\tau_1+a\tau_2)\quad,\quad\phi_3=c\tau_3
\]
with arbitrary numbers $a,b,c$ (the factors of $2^{-\frac{1}{2}}$ are
introduced for the sake of normalization). The conjugate momenta take
the form
\[
  p^1=2^{-\frac{1}{2}}(p_a\tau_1+p_b\tau_2)\quad,\quad
  p^2=2^{-\frac{1}{2}}(-p_b\tau_1+p_a\tau_2)\quad,\quad p^3=p_c\tau_3\,.
\]
The symplectic structure is given by
\[
  \{a,p_a\}=\{b,p_b\}=\{c,p_c\}=\kappa\iota'
\]
and vanishing in all other cases.

In the case of isotropic models we have only the two homomorphisms
$\lambda_0\colon SU(2)\to SU(2),f\mapsto 1$ and $\lambda_1=\id$
(again, this will be called $\tau_3$-gauge; for ease of notation we
use the same letters for the homomorphisms as in the LRS case, which
is justified by the fact that the LRS homomorphisms are restrictions
of those appearing here). The Higgs equation takes the form
$\epsilon_{iIK}\phi^j_K=0$ for $\lambda_0$ with no non-trivial
solution, and $\epsilon_{iIK}\phi^j_K=\epsilon_{ilj}\phi^l_I$. Each of
the last equations has the same form as for LRS models with $k=1$,
and their solution is $\phi^i_I=c\delta^i_I$ with an arbitrary
$c$.  In this case the conjugate momenta can be written as
$p^I_i=p\delta^I_i$, and we have the symplectic structure
$\{c,p\}=\kappa\iota'$.

Thus, we see that in both cases there is a unique non-trivial sector,
and no topological charge appears.

To be more concrete we calculate now the form of metric tensors for
Bianchi I and IX, and its related isotropic models. Because we have
the dreibein components $e^i_I$ which is the inverse matrix of
$p^I_i$, line elements are given by $\md
s^2=e^i_Ie_{iJ}\omega^I\otimes\omega^J$. Thus we have to calculate the
left invariant one-forms on $\R^3$ for Bianchi I, and on $SU(2)$ for
Bianchi IX. For Bianchi I we clearly have $\omega^I=\md x^I$ in a
coordinate system adapted to the translational symmetry. The line
element is $\md s^2=e^i_Ie_{iJ}\md x^I\otimes\md x^J$, and in case of
isotropy, i.e.\ $e^i_I=e\delta^i_I$, $\md s^2=e^2[(\md x^1)^2+(\md
x^2)^2+(\md x^3)^2]$ which is the metric of an isotropic flat
universe.

For Bianchi IX, i.e.\ for the translational symmetry group $SU(2)$,
the left invariant one-forms can be calculated, e.g.\ by using the
parameterization $g=\exp(2rn^i\tau_i)\in SU(2)$ with $0\leq r<2\pi$
and
$n=(\sin\vartheta\cos\varphi,\sin\vartheta\sin\varphi,\cos\vartheta)$,
from $g^{-1}\md g=\omega^I\tau_I$:
\begin{eqnarray*}
 \omega^1 & = & 2\sin\vartheta\cos\varphi\md r+ [\sin
 2r\cos\vartheta\cos\varphi- (\cos 2r-1)\sin\varphi]\md\vartheta\\
 & & +[-\sin 2r\sin\vartheta\sin\varphi- (\cos
 2r-1)\sin\vartheta\cos\vartheta\cos\varphi]\md\varphi\\
 \omega^2 & = & 2\sin\vartheta\sin\varphi\md r+ [\sin
 2r\cos\vartheta\sin\varphi+ (\cos 2r-1)\cos\varphi]\md\vartheta\\
 & & +[\sin 2r\sin\vartheta\cos\varphi- (\cos
 2r-1)\sin\vartheta\cos\vartheta\sin\varphi]\md\varphi\\
 \omega^3 & = & 2\cos\vartheta\md r- \sin 2r\sin\vartheta\md\vartheta+
 (\cos 2r-1)\sin^2\vartheta\md\varphi\,.
\end{eqnarray*}
In the isotropic case we obtain the metric
\[
 \md s^2=e^2\sum_{I=1}^3\omega^I\otimes\omega^I= 4e^2[(\md
 r)^2+\sin^2r\md\Omega^2]
\]
of an isotropic closed model of positive spatial curvature.

Finally, we have to specialize the constraints derived in the preceding
subsection to the enhanced symmetry. To that end we first relax the
partial gauge fixing introduced above by choosing the representative
$\lambda_1$. In general we can choose any homomorphism out of its
conjugacy class, i.e.\ $\lambda_1$ can be replaced by $g^{-1}\lambda_1
g$ for any $g\in G=SU(2)$. For the LRS models this amounts to
replacing $\tau_i$ in the expressions for $\phi_I$ by $g^{-1}\tau_i
g=:\Lambda^j_i\tau_j$. Thus, we obtain
\[
 \phi^i_1 = 2^{-\frac{1}{2}}(a\Lambda^i_1+b\Lambda^i_2)\quad,\quad
 \phi^i_2 = 2^{-\frac{1}{2}}(-b\Lambda^i_1+a\Lambda^i_2)\quad,\quad
 \phi^i_3 = c\Lambda^i_3
\]
and analogously for $p^I_i$. The matrix $\Lambda$ fulfills the
relations $\Lambda^k_i\Lambda^j_k=\delta^j_i$ and
$\epsilon_{ijk}\Lambda^i_l\Lambda^j_m\Lambda^k_n=\epsilon_{lmn}$,
which can be derived by calculating $\tr(g^{-1}\tau_igg^{-1}\tau^jg)$
and $\tr(g^{-1}\tau_lgg^{-1}\tau_mgg^{-1}\tau_ng)$, respectively.  The
expressions for isotropic models can be obtained by setting $b=0$ and
$a=\sqrt{2}c$.

The Gau\ss\ constraint now takes the form
\begin{eqnarray}
  {\cal G}^i\!\!\! & = &\!\!\! (\kappa\iota')^{-1}
  \epsilon_{ijk}\left[cp_c\Lambda^j_3\Lambda^k_3+
  {\textstyle\frac{1}{2}}(a\Lambda^j_1+b\Lambda^j_2) 
  (p_a\Lambda^k_1+p_b\Lambda^k_2)+
  \textstyle{\frac{1}{2}}(-b\Lambda^j_1+a\Lambda^j_2)
  (-p_b\Lambda^k_1+p_a\Lambda^k_2)\right]\nonumber\\
  & = &\!\!\! (\kappa\iota')^{-1}(ap_b-bp_a)\Lambda^i_3\,.\label{gausslrs}
\end{eqnarray}
To simplify the diffeomorphism constraint we use the fact that for
class A models the structure constants can be written as
$c^K_{IJ}=\epsilon_{IJL}n^{LK}$ with a symmetric matrix $n^{LK}$ which
can be diagonalized to $n^{LK}=n^{(K)}\delta^{LK}$ with eigenvalues
$n^{(K)}$. The (dedensitized) constraint then becomes
\begin{equation}\label{diffLRS}
  {\cal D}_aN^a = -(2\kappa\iota')^{-1}N^3
  (-c^1_{32}+c^2_{31})(ap_b-bp_a)= -{\textstyle\frac{1}{2}}
  (n^{(1)}+n^{(2)})(\kappa\iota')^{-1} N^3(ap_b-bp_a)\,,
\end{equation}
which vanishes already if the Gau\ss\ constraint is solved. This is a
consequence of an interrelation of $SU(2)$-gauge transformations and
diffeomorphisms due to the Higgs constraint which will be explained
in more detail when quantizing the constraints.

The Euclidean part of the dedensitized Hamiltonian constraint is
\begin{eqnarray}
  \frac{{\cal H}^{(E)}}{g_0} & = & -(n^{(1)}+n^{(2)})(ap_a+bp_b)p_c-
  n^{(3)}c(p_a^2+p_b^2)\nonumber\\
  & & +(ap_a+bp_b+cp_c)^2-{\textstyle\frac{1}{2}}(ap_a+bp_b)^2-(cp_c)^2+
  {\textstyle\frac{1}{2}}(ap_b-bp_a)^2\,.\label{hamlrs}
\end{eqnarray}

For isotropic models ($a=b=c$, $p_a=p_b=p_c=:p$) the constraints
${\cal G}^i$ and ${\cal D}_a$ vanish identically, whereas the
Euclidean part of the Hamiltonian constraint takes the form
\begin{equation}\label{hamisotropic}
  \frac{{\cal H}^{(E)}}{g_0}= -2(n^{(1)}+n^{(2)}+n^{(3)})c p^2+6(c
  p)^2
\end{equation}
where $n^{(1)}=n^{(2)}=n^{(3)}=0$ for isotropic flat (Bianchi I), and
$n^{(1)}=n^{(2)}=n^{(3)}=1$ for isotropic closed (Bianchi IX).

\subsection{Auxiliary Hilbert Spaces for Homogeneous Models}

The configuration spaces for our models are given by Higgs `fields' in
a single point, which shows that they are finite-dimensional. In
quantum theory they will be represented as spaces of point holonomies
\cite{FermionHiggs} associated to a single point, the only point $x_0$
in the reduced manifold $B$. For the anisotropic models with an empty
Higgs condition (\ref{Higgs}) there are three independent
$SU(2)$-Higgs fields $\phi_1$, $\phi_2$ and $\phi_3$ associated to the
independent directions $T_I$ in the tangent space of $x_0$. Thus, we
have three point holonomies $h_I:=\exp\phi^i_I\tau_i$ lying in a
single vertex (the point $x_0$), in which $SU(2)$-gauge invariance has
to be imposed. The auxiliary Hilbert space on which the constraints
have to be solved is the space $\Haux=L^2([SU(2)]^3,[\md\mu_H]^3)$ of
functions of the three point holonomies. Its measure is analogous to
the Ashtekar--Lewandowski measure ($\md\mu_H$ is the Haar measure on
$SU(2)$). The momenta $p^I_i$ will be represented as derivative
operators on functions in $\Haux$. To calculate their action we need
the small

\begin{lemma}
  Let $G$ be a Lie group and $F\colon\R\to{\cal L}G$ be a
  differentiable ${\cal L}G$-valued function in a real parameter. The
  derivative of $\exp F(s)\in G$ with respect to $s$ in the
  point $s=s_0$ is given by
  \[
    \left.\frac{\md\exp F(s)}{\md s}\right|_{s=s_0}=\int_0^1\md
    t\exp(tF(s_0))F'(s_0)\exp((1-t)F(s_0))
  \]
  where $F'$ is the derivative of $F$ with respect to $s$.
\end{lemma}

\begin{proof}
  The derivative
\[
  \left.\frac{\md\exp F(s)}{\md s}\right|_{s=s_0}\!\!\!\!\!:=
  \lim_{s\to s_0} \frac{\exp F(s)-\exp
    F(s_0)}{s-s_0}= \lim_{s\to s_0} \frac{\exp
    F(s)\exp(-F(s_0))-1}{s-s_0}\exp F(s_0)
\]
  can be written as
\begin{eqnarray*}
  \left.\frac{\md\exp F(s)}{\md s}\right|_{s=s_0} & = & 
  \lim_{s\to s_0} (s-s_0)^{-1}\int_0^1\md t\frac{\md}{\md t}[\exp
  (tF(s)) \exp(-tF(s_0))]\exp F(s_0)\\
  & = & \lim_{s\to s_0}\int_0^1\md t \exp (tF(s))
  \frac{F(s)-F(s_0)}{s-s_0} \exp(-tF(s_0))\exp F(s_0)\\
  & = & \int_0^1\md t\exp(tF(s_0))F'(s_0)\exp((1-t)F(s_0))\,.
\end{eqnarray*}
\end{proof}

Applied to $\exp\phi^i_I\tau_i$ we get for the action of
$\frac{\partial}{\partial \phi^j_J}$
\[
  \frac{\partial}{\partial \phi^j_J}\exp\phi^i_I\tau_i=
   \delta^J_I\int_0^1\md t
  \exp(t\phi^i_I\tau_i) \tau_j \exp((1-t)\phi^i_I\tau_i)
\]
which cannot be represented as an element of the auxiliary Hilbert
space (it contains a continuous family of point holonomies).
Analogously to \cite{FermionHiggs} we can regularize the Higgs vertex
by smearing the point holonomy to a holonomy associated with a
regularizing edge. This introduces a $\delta$-function into the action
of $\frac{\partial}{\partial \phi^i_I}$ which is non-vanishing only in
the endpoints of the edge corresponding to $t=0$ and $t=1$ in the
formula of the lemma, and the momenta get quantized to combinations of
left and right invariant vector fields
\begin{eqnarray}
  \hat{p}^I_i & = & -i\iota'l_P^2\sum_{J=1}^3 
  \frac{\partial (h_J)^A_B}{\partial\phi^i_I(x_0)}
  \frac{\partial}{\partial (h_J)^A_B} =-i\iota'l_P^2\sum_{J=1}^3
    {\textstyle\frac{1}{2}} (\tau_i
  h_I+h_I\tau_i)^A_B \frac{\partial}{\partial (h_I)^A_B}\nonumber\\
  & = & -{\textstyle\frac{1}{2}}i\iota'l_P^2\left(X_i^{(L)}(h_I)+
  X^{(R)}_i(h_I)\right)
\end{eqnarray}
acting on the $I$-th copy of $SU(2)$ in the domain of definition of a
function in $\Haux$ (the $\delta$-function is integrated to
$\frac{1}{2}$ because its singularity lies at the endpoints of the
domain of integration).

Of course, in $B$ there is no place for a regularizing edge.
Therefore, we introduce a compact auxiliary manifold homeomorphic to
$\overline{S/F}$. The bar reminds us that we may have to take a
certain compactification (e.g.\ the one-point compactification of
$\R^3$ for Bianchi I) of $S/F$ (this is not necessary for Bianchi IX),
and we take into account the possibility of a non-trivial isotropy
subgroup $F$ for later use. We need this space only as a
differentiable manifold: The group structure of $S$ does not play any
role here. On $S/F$ we have the vector fields $X_I$ dual to the left
invariant one-forms $\omega^I$ used earlier. With their help we define
curves $e_I\colon[0,1]\to\overline{S/F}$ by the differential equation
$\dot{e}_I(t)=X_I(e_I(t))$, and we assume the compactification of
$S/F$ to be taken in such a way that the three curves $e_I$ are
closed. We take them as regularizing edges for the three point
holonomies, which is justified
by the equation
\[
  h(e_I) := {\cal P}\exp\int_0^1\md t\dot{e}^a_IA^i_a\tau_i= {\cal
    P}\exp\int_0^1\md t\phi^i_J\omega^J(\dot{e}_I)\tau_i=
  \exp(\phi^i_I\tau_i)
\]
for the holonomy along $e_I$ of a reconstructed connection on
$\overline{S/F}$. The auxiliary Hilbert space is then generated by
spin network states associated with graphs consisting of the three
closed edges $e_I$, which meet in the $6$-vertex $x_0$, and which are
labeled by spins $j_I\in\frac{1}{2}\N_0$. There is some arbitrariness
in the directions of the $e_I$: Depending on the special model there
is gauge freedom due to the diffeomorphism constraint which acts by
inner automorphisms of $S$. This freedom will be fixed by solving the
diffeomorphism constraint in Section~\ref{s:diff}, for the moment we
can choose some appropriate directions for the $e_I$ which amounts to a
total gauge fixing of reduced diffeomorphisms.

For a non-trivial isotropy subgroup $F$ the situation is more
complicated. At first, classically the gauge is fixed partially: For
LRS models the reduced gauge group is $U(1)$, whereas for isotropic
models it is fixed totally and there is no gauge group at all. These
facts are nicely illustrated by the Gau\ss\ constraint for these
models, which is proportional to $\Lambda^i_3$, which forms the
internal axis of the remaining gauge freedom for LRS models with gauge
fixing given by $\Lambda$, or vanishes completely in case of isotropic
models. In the quantum theory the partial gauge fixing can (and has
to) be undone (here lies the advantage of the general framework
described in Section~\ref{s:symmred}). Thus, we will use
$SU(2)$-holonomies also for LRS and isotropic models.

In addition, one has the Higgs equation (\ref{Higgs}) to be solved. By
exponentiating it can be written as
\begin{equation}\label{HiggsHol}
  h(f(e_I)):=\exp\phi(\Ad_f(T_I))=\exp\Ad_{\lambda(f)}\phi(T_I)=
  \Ad_{\lambda(f)}\exp\phi(T_I)=\Ad_{\lambda(f)}h(e_I)
\end{equation}
and interpreted as a condition for holonomies which are obtained by
rotating the edges $e_I$. As a first application of this equation,
which provides a geometrical interpretation of the Higgs constraint,
one can easily see that usage of the homomorphism $\lambda_0$ does not
lead to a non-vanishing Higgs field: The right hand side is then
identically $h(e_I)$, whereas for $f$ a rotation by $180^o$ the left
hand side becomes $h(e_I)^{-1}$. The equation can be fulfilled only
for $h(e_I)=1$, i.e.\ a vanishing Higgs field. This consideration can be
extended to all even values of $k$. Furthermore, we see that rotated
holonomies are gauge equivalent, and therefore some of them are
redundant. Roughly, this leads to only two independent holonomies for
LRS models (an axial one $h(e_3)$, which we choose in $3$-direction,
and a transversal one $h(e_1)$ representing the two equivalent
holonomies), and only one for isotropic models. However, there are
subtleties because of the twisting introduced by gauge rotations on
the right hand side of equation (\ref{HiggsHol}). To make it clear we
first treat the $\lambda_0$-sectors pretending, for illustrative
purposes only, that they would lead to non-trivial Higgs fields.

For anisotropic models the classical configuration space is given by
${\cal U}=SU(2)^3=\{(\exp(\phi_1^i\tau_i), \exp(\phi^i_2\tau_i),
\exp(\phi^i_3\tau_i))\}$.  Equation (\ref{HiggsHol}) then states that
all three holonomies are the same for an isotropic model, i.e.\ in
this case we have the configuration space ${\cal
  U}^{[\lambda_0]}_{\mathrm{iso}}=\{(h,h,h):h\in SU(2)\}$, which is
the diagonal $SU(2)$-subgroup of $SU(2)^3$. This is the would-be
solution space of the Higgs condition leading to the auxiliary Hilbert
space $L^2(SU(2),\md\mu_H)$ spanned by spin networks associated with
graphs consisting of a single closed edge. In the case of LRS models
this consideration leads to two independent holonomies. But the
situation described in the present paragraph does not appear, because
for $\lambda_0$ we have no non-trivial Higgs field. Therefore, we have
to determine the classical configuration space for the realistic
$\lambda_1$-sectors with their twisting in equation (\ref{HiggsHol}).

For LRS models in the $\tau_3$-gauge, we have again a subspace of
$SU(2)^3$ parameterized by the parameters $a=:A\cos\alpha$,
$b=:A\sin\alpha$, $c$ introduced above as
\[
  {\cal U}^{\tau_3}_{\mathrm{LRS}}=\{(\exp (A\,n^i(\alpha)\tau_i), \exp
  (A\,\epsilon_{3ij}n^i(\alpha)\tau^j), \exp (c\tau_3))\}
\]
with $n^i=(\cos\alpha,\sin\alpha,0)$. The parameter $\alpha$ is pure
gauge, whereas the parameters $A$, $c$ represent the gauge invariant
information. The configuration space on which the partial gauge fixing
is undone is the union of the conjugacy classes of all elements of
${\cal U}^{\tau_3}_{\mathrm{LRS}}$. It can be written as
\begin{equation}\label{Ulambda}
  {\cal U}^{[\lambda_1]}_{\mathrm{LRS}}=\{(\exp (A\Lambda_1^i\tau_i),\exp
  (A\Lambda_2^i\tau_i), \exp (c\Lambda^i_3\tau_i))\}
\end{equation}
and depends only on the conjugacy class $[\lambda_1]$. It is
parameterized by five parameters: $A$, $c$, and the three angles which
determine the dreibein $\Lambda^i_j$. For isotropic models we can
obtain ${\cal U}^{[\lambda_1]}_{\mathrm{iso}}$ by setting $A=c$.

Going over from ${\cal U}^{\tau_3}$ to ${\cal U}^{[\lambda_1]}$
introduces no new degrees of freedom as long as we require functions
on ${\cal U}^{[\lambda_1]}$ to be gauge invariant under diagonal
$SU(2)$-conjugation. This restores the original gauge group $SU(2)$ of
the non-symmetric theory. Our reason for doing so is two-fold: First,
the partial gauge fixing is undone, and the reduced theory will only
depend on $[\lambda_1]$, not on the selection of a representative.
Second, we will be able to use techniques developed for $SU(2)$-spin
networks, which would not be possible in the gauge fixed case.

Now we can see the difference between the fake $\lambda_0$-case above
and the realistic $\lambda_1$-case: For $\lambda_0$ the solution space
of the Higgs equation was a subgroup of $SU(2)^3$. Thus we could use
the Peter--Weyl theorem to identify all functions on this manifold
with matrix elements of $SU(2)$-representations, which lead us to spin
networks associated with a reduced number of edges. However, due to
the twisting for $\lambda_1$ the solution space is no longer a
subgroup, but only a union of conjugacy classes in $SU(2)^3$. The
Peter--Weyl theorem does no longer apply, and we have to determine all
functions on ${\cal U}^{[\lambda_1]}$ by hand.  Of course, the spin
network states with a reduced number of edges are some of those
functions. E.g., for isotropic models they can be obtained from spin
networks in the anisotropic theory with two of the three labelings
being zero. But these do not comprise all functions on ${\cal
  U}^{[\lambda_1]}_{\mathrm{iso}}$: One can easily see that all such
gauge invariant spin network functions with one edge are symmetric
under $c\to-c$ if they are evaluated in $h=\exp (c\Lambda_3)$.  But
there are gauge invariant functions on ${\cal
  U}^{[\lambda]}_{\mathrm{iso}}$ which are not symmetric under this
reflection: One example is given by $\tr[\exp (c\Lambda_1)\exp
(c\Lambda_2)\exp (c\Lambda_3)]= 2\cos^3(c/2)-2\sin^3(c/2)$, which
stems from an anisotropic spin network with all labelings being
$\frac{1}{2}$ and an appropriate vertex contractor. The situation for
LRS models is similar: Ordinary spin networks with two edges do not
suffice.

Thus there are more gauge invariant functions on the classical
configuration space than naively expected. We will discuss this in
more detail and derive all such functions in the second part
\cite{cosmoII}. Regarding the purposes of the present part, i.e.\
quantization and solution of Gau\ss\ and diffeomorphism constraints,
no important new features are introduced by these additional
functions. At this stage it suffices to know that all such functions
can be viewed as spin network functions with two (LRS) or one
(isotropic) edge, but possibly with a certain insertion in the vertex
$x_0$ (for isotropic models the auxiliary Hilbert space is doubled in
this way) which does neither affect the action of gauge
transformations nor of diffeomorphisms. This insertion can be viewed
as a reduction of the vertex contractor in spin networks with three
closed edges, which was not taken into account in the naive arguments
presented above.

\section{Gau\ss\ Constraints}
\label{s:gauss}

We proceed now by quantizing the Gau\ss\ constraint
(\ref{gaussbianchi}) on the auxiliary Hilbert space derived in the
preceding section. To that end we introduce new parameters which substitute
the Higgs field components $\phi^i_I$ and which are better suited for
this purpose. These new coordinates $r_I$, $\alpha^3_I$, $\beta^3_I$
are, independently for each $1\leq I\leq 3$, defined by
$\phi^i_I=:r_In_3^i(\alpha^3_I,\beta^3_I)$ with
$n_3^i(\alpha,\beta):=(\cos\alpha\sin\beta,\sin\alpha\sin\beta,\cos\beta)$.
Alternatively, we can choose the parameters $\alpha^2_I$, $\beta^2_I$
defined analogously with
$n_2^i(\alpha,\beta):=(\sin\alpha\sin\beta,\cos\beta,\cos\alpha\sin\beta)$,
or the parameters $\alpha^1_I$, $\beta^1_I$ with
$n_1^i(\alpha,\beta):=(\cos\beta,\cos\alpha\sin\beta,\sin\alpha\sin\beta)$.
It is easy to see that these new parameters fulfill
\[
  \frac{\partial}{\partial \alpha^i_I}=\epsilon_{ijk}\phi^j_I
   \frac{\partial}{\partial \phi^k_I} \quad \mbox{(no
    sum over $I$)}\:.
\]
Thus,
we can quantize the Gau\ss\ constraint (\ref{gaussbianchi}) acting on 
a function $f\in\Haux$ as
\[
 \hat{{\cal G}}_i f=\frac{\hbar}{i} \epsilon_{ijk}
 \phi^i_I\frac{\partial f}{\partial\phi^k_I}= \frac{\hbar}{i}
 \sum_{I=1}^3\frac{\partial f}{\partial\alpha^i_I}
\]
where we use either of the three sets of angular parameters depending
on the component of the Gau\ss\ constraint.

To calculate the derivative we note that a point holonomy
\[
  h_I=\exp(\phi^i_I\tau_i)=\exp (r_In^i_3(\alpha^3_I,\beta^3_I)\tau_i)
\]
gets changed under a gauge transformation with $\exp(\gamma\tau_3)$ into
\[
  \exp(\gamma\tau_3) \exp(r_In^i_3(\alpha^3_I,\beta^3_I)\tau_i)
  \exp(-\gamma\tau_3)=
  \exp (r_In^i_3(\alpha^3_I+\gamma,\beta^3_I)\tau_i)\,.
\]
Thus $g_I$ can be
written as $h_I=\exp(\alpha^3_I\tau_3)
\exp(r_In^i_3(0,\beta^3_I)\tau_i) \exp(-\alpha^3_I\tau_3)$, and
analogously for $\alpha^2_I$ or $\alpha^1_I$. The
derivative of $h_I$ with respect to $\alpha^i_I$ can now be calculated
as
\[
  \frac{\partial h_I}{\partial\alpha^i_I}=\tau_ih_I-h_I\tau_i\,.
\]
With this relation the Gau\ss\ constraint acting on the function
$f(h_1,h_2,h_3)$ becomes
\begin{eqnarray*}
  \hat{{\cal G}}_i f & = & \frac{\hbar}{i} \sum_{I=1}^3
  \frac{\partial f}{\partial\alpha^i_I}= \frac{\hbar}{i} \sum_{I=1}^3
  \left(\frac{\partial (h_I)^A_B}{\partial\alpha^i_I}\right) 
  \frac{\partial f}{\partial (h_I)^A_B}\\
  & = & \frac{\hbar}{i}
  \sum_{I=1}^3(\tau_ih_I-h_I\tau_i)^A_B 
   \frac{\partial f}{\partial (h_I)^A_B}= 
  \frac{\hbar}{i} \sum_{I=1}^3
  \left(X_i^{(R)}(h_I)-X_i^{(L)}(h_I)\right) f
\end{eqnarray*}
with the difference of a right and left invariant vector field for
each point holonomy. This action is as expected, because each of the
point holonomies transforms with respect to the adjoint
representation under a gauge transformation.

The solution of the constraint can be given in a standard way by
restricting the Hilbert space to the gauge invariant subspace spanned
by gauge invariant spin network states, i.e.\ those spin networks
whose intertwiner in the vertex $x_0$ contracts the six
representations (an incoming and an outgoing for each of the three
edges) to the trivial representation.

For LRS models we have to quantize the reduced constraint
(\ref{gausslrs}). After introducing the angle $\alpha:=\arctan b/a$ it
can, analogously to the calculations above, be written as
\[
  \hat{{\cal G}}_if= \frac{\hbar}{i}
  \Lambda_3^i\,\Lambda_3^j\left(X_j^{(R)}(h_1)-X_j^{(L)}(h_1)\right)f\,.
\]
Note that after solving the Higgs constraint a function $f$ in the
auxiliary Hilbert space only depends on the two point holonomies
$h_1=\exp(a\Lambda^i_1+b\Lambda^i_2)\tau_i$ and $h_3=\exp
(c\Lambda^i_3\tau_i)$.

There are two points to mention about this operator. First,
classically $\Lambda^3_i$ is fixed so that the quantization of the
Gau\ss\ constraint consists of only one component of invariant vector
fields ($X_3$ in the $\tau_3$-gauge), and it would force only this
component to vanish if we would use this partial gauge fixing in
quantum theory. This corresponds to the fact that the reduced
structure group for $F=U(1)$ is $U(1)$ consisting of internal
rotations around an axis determined by the partial gauge fixing
$\Lambda_3^i$. However, we already relaxed the partial gauge fixing to
arrive at our auxiliary Hilbert space. In this process functions on
the partially gauge fixed configuration space ${\cal
  U}^{\tau_3}_{\mathrm{LRS}}$ can be extended to functions on ${\cal
  U}^{[\lambda_1]}_{\mathrm{LRS}}$ by demanding invariance under
conjugation. Only for those functions the gauge fixed Gau\ss\ 
constraint can be used. On an arbitrary function on ${\cal
  U}^{[\lambda_1]}_{\mathrm{LRS}}$ we have to impose all three
components of an $SU(2)$-constraint. (For a similar discussion in case
of spherically symmetric quantum gravity see reference
\cite{SymmRed}.)

Second, the operator contains only vector fields associated with the
holonomy $h_1$, whereas the axial holonomy $h_3$ does not appear at
all. At first one would expect both holonomies to contribute equally,
because the auxiliary Hilbert space is spanned by spin network states
with the two edges $e_1$ and $e_3$ meeting in the vertex $x_0$. But
after taking into account the construction of the solution space
${\cal U}^{[\lambda_1]}_{\mathrm{LRS}}$ of the Higgs equation,
vanishing of the $e_3$-contribution is completely consistent: The
Gauss constraint contains the $\Lambda_3^i$-component of vector
fields, and due to $h_3=\exp (c\Lambda^i_3\tau_i)$ on ${\cal
  U}^{[\lambda_1]}_{\mathrm{LRS}}$ ($\Lambda$ is now a
coordinate on that space) we have
\[
  \Lambda^i_3X_i^{(R)}(h_3)= \tr\left[(
    \Lambda^i_3\tau_ih_3)^T\frac{\partial}{\partial h_3}\right]= \tr\left[
    (h_3\Lambda^i_3\tau_i)^T \frac{\partial}{\partial h_3}\right]=
  \Lambda^i_3X_i^{(L)}(h_3)\,.
\]
Thus, the $\Lambda^i_3$-components of the right and left invariant
vector fields are the same, and they cancel one another in the Gau\ss\ 
constraint. Therefore, they do not appear anymore in the partially
fixed Gau\ss\ constraint acting on functions on ${\cal
  U}^{[\lambda_1]}_{\mathrm{LRS}}$.

After this discussion we see that the Gau\ss\ constraint can be
solved in the quantum theory, without partial gauge fixing, by gauge
invariant $SU(2)$-spin networks. Here, only two edges meet in the
$4$-vertex $x_0$ (the insertion mentioned in the preceding section
does not affect this consideration).

Having the discussion of LRS models in mind we can treat the isotropic
briefly. Classically, the gauge group is completely broken,
$Z_G(\lambda_1(F))=\{1\}$, and one would not expect a Gau\ss\ 
constraint. However, in quantum theory after undoing the gauge fixing
we use $SU(2)$-spin networks with a single edge, which should be gauge
invariant in the vertex $x_0$, i.e.\ the two copies of the
representation to the label $j$, one for the incoming and one for the
outgoing part of the closed edge, should be contracted to the trivial
representation in $x_0$.

\section{Diffeomorphism Constraints}
\label{s:diff}

Before quantizing the diffeomorphism constraint we describe shortly
the role played by diffeomorphisms in symmetry reduced models. By using
ans\"atze for invariant fields adapted to the symmetry some freedom in
applying diffeomorphisms is fixed. Classically, this arises because
one uses special coordinates which exhibit the symmetry, e.g.\ polar
coordinates in case of spherical symmetry. Therefore, only
diffeomorphisms respecting the ans\"atze are realized in the symmetry
reduced theory. These are typically diffeomorphisms of the reduced
manifold $B$, e.g.\ a radial manifold in case of spherical symmetry,
whereas the remaining freedom is fixed. But for homogeneous models
studied here the reduced manifold $B=\{x_0\}$ consists of a single
point, and one may ask why the diffeomorphism constraint
(\ref{diffbianchi}) does not vanish, for there are now no
diffeomorphisms of the reduced manifold.

A hint for an answer to that question comes from the fact that
(\ref{diffbianchi}) generates inner automorphisms of the symmetry
group $S$, the group manifold of which is identified (modulo
compactification) with the homogeneous space manifold $\Sigma$. This
can be seen from the fact that the expression (\ref{diffbianchi}) is
similar to the constraint (\ref{gaussbianchi}), which generates
conjugation in the internal space, except for an exchange of the
structure constants $\epsilon_{ijk}$ of $SU(2)$ with the ones
$c^K_{IJ}$ of $S$.

The remaining freedom after choosing coordinates adapted to the
symmetry is to select an origin $x_0$ of the coordinate system.
Instead of $x_0$ we could choose any other point $sx_0$, $s\in S$ in
$\Sigma$ (we can indeed reach any other point owing to transitivity of
the group action). Using the base point $x_0$ all points $gx_0$ in
$\Sigma$ can be parameterized by the group coordinates of $g$ (this
group element is unique if there is no isotropy subgroup, otherwise we
can use coordinates of $gF_{x_0}$ in the homogeneous space
$S/F_{x_0}$). But after performing a left translation with $s$, which
shifts the origin to $sx_0$, the point $gx_0$ is mapped to
$sgx_0=(sgs^{-1})sx_0$. Thus upon changing the base point the role of
$S$ (providing coordinates on $\Sigma$) is played by the isomorphic
group $sSs^{-1}$. Inner automorphisms of $S$ are the remaining gauge
freedom under the diffeomorphism group, and the diffeomorphism
constraint, which demands independence of the physical phase space
under inner automorphisms, can be seen to enforce independence of the
selection of a base point.  In the classically reduced manifold
(consisting of a single point) these transformations have, of course,
no geometric meaning. But in the course of quantization we introduced
an auxiliary manifold when promoting point holonomies to holonomies
associated with edges therein. The inner automorphisms act on this
manifold and thereby move these edges depending on the symmetry group
$S$.  Strictly speaking, we have to relax the gauge fixing of the
diffeomorphism group which we introduced implicitly earlier by fixing
three edges in the auxiliary manifold on which a spin network function
depends. To study the action of diffeomorphisms and to eventually
solve the constraint we have to allow spin networks associated to
graphs with three edges which can be transformed against the original
edges.

E.g., for Bianchi I all inner automorphisms are trivial and there is
no non-trivial action on the edges (the diffeomorphism constraint
vanishes in this case), and for Bianchi IX the group of inner
automorphisms is isomorphic to $SO(3)$ rotating the three edges. The
last example will be discussed below, because it will serve us to
discuss the difference between gauge and symmetry. Note that the
diffeomorphism group in this case acts identically to the additional
symmetry group ($F=SU(2)$) imposed when constraining Bianchi IX to an
isotropic model. But the treatment of gauge in the one case (by group
averaging) and of symmetry in the other (by quantum symmetry
reduction) is very different, as will be illustrated by this example.

\subsection{Quantization}

We now know the action which is generated by the diffeomorphism
constraint, and we can use it to solve the constraint by
group averaging. But for illustrative purposes we will first
investigate whether the constraint can be quantized in its
infinitesimal version. To that end we write the action of an inner
automorphism generated by $T_K\in{\cal L}S$ on a point holonomy as
\[
  \Ad(\exp(-\delta T_K))\colon S\to S,h\mapsto\exp(-\delta
  T_K)h\exp(\delta T_K)
\]
with a parameter $\delta\in\R$. By differentiation this determines a
map on the Lie algebra of $S$:
\begin{eqnarray*}
  \Ad(\exp(-\delta T_K))\colon{\cal L}S\to{\cal L}S,c^IT_I \mapsto
  \exp(-\delta T_K)c^IT_I\exp(\delta
  T_K) & = & c^I\exp\left(\delta(c^J_{IK})^J_I\right)T_J\\
  & =: & c^I\Ad^J_I(\exp(-\delta T_K))T_J
\end{eqnarray*}
where the matrix elements $\Ad^J_I(\exp(-\delta T_K))$ are defined in
terms of the matrix exponential of the matrix $(c^J_{IK})^J_I$.

Because the edges $e_I\colon[0,1]\to\overline{S/F}$ are defined by its
direction $T_I$ in the identity of $S$, they get transformed into
\begin{eqnarray*}
  e_I(t) & \mapsto & e_I^{(\delta,K)}(t):=\exp(-\delta T_K)\exp (tT_I)
  \exp(\delta T_K)\:,\\
  & & \dot{e}^{(\delta,K)}_I(t)=\Ad^J_I(\exp(-\delta T_K))\dot{e}_J(t)
\end{eqnarray*}
which is an integral curve to the left invariant vector field
$\Ad^J_I(\exp(-\delta T_K))X_J$. The holonomy along this new edge is
\begin{eqnarray*}
  h(e_I^{(\delta,K)}) & = & {\cal P}\exp\int_0^1\md t\,
  \phi^i_J\omega^J(\Ad^L_I(\exp(-\delta T_K))X_L)\tau_i\\
 & = & {\cal
    P}\exp\int_0^1 \md t\, \phi^i_L\Ad^L_I(\exp(-\delta T_K))\tau_i\\
 & = & {\cal P}\exp\int_0^1 \md t\, \phi^i_L(\delta^L_I+\delta c^L_{IK}+
 O(\delta^2))\tau_i= \exp[(\phi^i_I+\delta c^L_{IK}\phi^i_L+
 O(\delta^2))\tau_i]\\
 & = & \exp(\phi^i_I\tau_i)+\delta\frac{\md}{\md \delta}
  \exp((\phi^i_I+\delta
 c^L_{IK}\phi^i_L+ O(\delta^2))\tau_i)|_{\delta=0}+O(\delta^2)\\
 & = & \exp(\phi^i_I\tau_i)+ \delta c^L_{MK}\phi^j_L
  \frac{\partial}{\partial \phi^j_M}
 \exp(\phi^i_I\tau_i)+ O(\delta^2)
\end{eqnarray*}
where we Taylor expanded in $\delta$. If we apply a function $f$ to the
transformed holonomy and again Taylor expand, we obtain
\begin{eqnarray*}
  f\left(\!h(e_I^{(\delta,K)})\!\right)-f(h(e_I))\!\!\! & = &\!\!\!
  f\left(\!\exp(\phi^i_I\tau_i)+ \delta
  c^L_{MK}\phi^j_L\frac{\partial}{\partial \phi^j_M}\exp(\phi^i_I\tau_i)
  +O(\delta^2)\!\right)-
  f(\exp(\phi^i_I\tau_i))\\
  & = &\!\!\! \delta c^L_{MK}\phi^j_L 
  \frac{\partial h(e_I)^A_B}{\partial\phi^j_M}
  \frac{\partial}{\partial h(e_I)^A_B} f(h(e_I))+O(\delta^2)\,.
\end{eqnarray*}
After replacing $p^I_i$ by a functional derivative with respect to
$\phi^i_I$ in the course of quantization this already provides the correct
expression for a quantization of (\ref{diffbianchi}). Up to
$O(\delta^2)$ we obtain
\[
  \hat{{\cal D}}_K f=-i\hbar\delta^{-1}\left(
  f\left(h(e_1^{(\delta,K)}),
  h(e_2^{(\delta,K)}), h(e_3^{(\delta,K)})\right)-
  f(h(e_1),h(e_2),h(e_3))\right)+
  O(\delta^2)
\]
where $f$ depends on three holonomies $h(e_I)$, $1\leq I\leq 3$. The
component ${\cal D}_K$ is defined by ${\cal D}_aN^a=:{\cal
  D}_KN^K$. We would get a quantization of the constraint if we
could perform the limit $\delta\to 0$ in the last equation. In such a
case the diffeomorphism constraint would just act as Lie derivative.
But we encounter here the same problem as for the diffeomorphism
constraint in the full theory (Appendix C of \cite{ALMMT}): In the
diffeomorphism invariant Ashtekar--Lewandowski inner product the
functions associated with a graph consisting of the edges $e_I$ on the
one hand and of the edges $e_I^{(\delta,K)}$ on the other are
orthogonal for all $\delta\not=0$, and the limit does not exist in the
associated topology.

\subsection{Group Averaging}

Instead of quantizing the infinitesimal constraint we can use the
known action of diffeomorphisms on the auxiliary manifold to solve the
constraint by group averaging \cite{ALMMT}. This is a simple procedure
because all graphs underlying cylindrical functions in the auxiliary
Hilbert space have at most three edges. Generically, this will bring
us back to the space of functions on holonomies to three fixed edges
used earlier.  But graph symmetries have to be taken properly into
account \cite{ALMMT,QSDIII}, which will be done now for the example of
Bianchi IX.

In this case the action generated by the diffeomorphism constraint
consists of all rotations in the auxiliary manifold. Thus, it is the
same as the action of the isotropy subgroup for an isotropic closed
model. We will see how these different concepts of gauge and symmetry
are implemented. To solve the diffeomorphism constraint by group
averaging we first have to determine an allowed basis for the space of
spin network states according to reference \cite{QSDIII}. Allowed
states are defined by summing over the index set of labels
\[
\Xi(I):=\{\xi:\mbox{there is a }\phi\in\Diff\mbox{ with
  }U(\phi)T_I=T_{\gamma(I),\xi}\}
\]
where $I$ denotes a multi-label consisting of the graph $\gamma(I)$
and further labelings $\xi$ for a spin network $T_I$. $\Diff$ is the
diffeomorphism group, here $SO(3)$, and $U$ its representation on the
space of spin network functions. Furthermore, $n(I):=|\Xi(I)|$ is the
number of elements of an orbit of the label $I$. An allowed basis is
built from functions which are symmetric (this should not be confused
with the symmetry group of the symmetry reduction) with respect to
graph symmetries:
\[
  T^S_I:=n(I)^{-\frac{1}{2}}\sum_{\xi\in\Xi(I)}T_{\gamma(I),\xi}\:,
\]
and their group averaging is
\[
  [T^S_I]:=n(I)^{-\frac{1}{2}}[T_I]:=n(I)^{-\frac{1}{2}}
  \sum_{\phi\in\Diff} T_{\phi(I)}
\]
which solves the diffeomorphism constraint. All other states of the
allowed basis are annihilated by group averaging.

In the case of Bianchi IX all spin networks are associated with graphs
consisting of three edges meeting in a $6$-vertex which can all be
obtained as rotations of a fixed dreibein. In general, spin network
functions associated with graphs which are rotated against one another are
orthogonal, the only exception being the case of graph symmetries. The
group of graph symmetries is here the permutation group $S_3$ on the
three edges. If we define the subgroup $\sigma(T)\leq S_3$ as the group
of label symmetries which fix not only the graph but the whole
labeling when acting on a spin network $T$, we can write the symmetric
states of the allowed basis as
\[
  T^S=\sqrt{\frac{|\sigma(T)|}{|S_3|}}
  \sum_{\phi\in S_3/\sigma(T)}U(\phi)T\,.
\]
There are three different cases: For $j_1=j_2=j_3$, i.e.\ identical
labels for all three edges, we have $\sigma(T)=S_3$ and $T^S=T$. The
case $j_1=j_2\not=j_3$ (and analogously $j_1\not=j_2=j_3$,
$j_1=j_3\not=j_2$), $\sigma(T)\cong S_2$ leads to
$T^S=3^{-\frac{1}{2}}(T+ T_{j_1\leftrightarrow
  j_3}+T_{j_2\leftrightarrow j_3})$, and finally
$j_1\not=j_2\not=j_3\not=j_1$, $\sigma(T)=\{1\}$ to
$T^S=6^{-\frac{1}{2}}\sum_{\pi\in S_3}T_{\pi}$ (the subscript
indicates the permutation performed on the edges and their labelings).
These are all independent states which survive group averaging. We see
that we essentially come back to the gauge fixed states with only
three fixed edges, the only novelty being implied by the
symmetrization with respect to $S_3$. But there are still the three
labels $j_1$, $j_2$ and $j_3$, and certainly the vertex contractor
which are all needed to specify a state. The symmetrization implies
only a minor decrease in the freedom. In contrast, if we treat $SO(3)$
as a symmetry group for an isotropic model, we have seen that there
remains only one edge labeled by a single spin $j$, and the insertion
mentioned earlier, which can be viewed as a remnant of the vertex
contractor. This illustrates the difference between the different
treatments of symmetry and gauge: Solving the gauge constraint
eliminates redundant degrees of freedom which are given by the ability
to choose an arbitrary dreibein to represent the edges (interpreted in
terms of the auxiliary manifold). The symmetry reduction reduces the
number of degrees of freedom even stronger by selecting particular
geometries, and therefore the number of spin network labels is
reduced.

The diffeomorphism constraint for isotropic models vanishes
identically which is consistent with the discussion above: Symmetry
reduction is stronger than gauge reduction, and therefore an isotropic
state is automatically invariant with respect to inner automorphisms.

For LRS models the situation is different: Here, the constraint
(\ref{diffLRS}) either vanishes identically ($n^{(1)}+n^{(2)}=0$) or
is equivalent to the Gau\ss\ constraint, i.e.\ it is already solved by
using gauge invariant states. This is a consequence of the Higgs
constraint, which can most easily be seen in the form
(\ref{HiggsHol}). For LRS models we have $n^{(1)}=n^{(2)}$ (see
reference \cite{MacCallum}), which is non-zero in the case of a
non-vanishing diffeomorphism constraint. In this case the only inner
automorphism with respect to which there is a non-vanishing component
of the diffeomorphism constraint is a rotation around the axial edge.
But owing to the Higgs constraint (\ref{HiggsHol}) such a rotation
applied to a transversal edge is equivalent to a gauge rotation of the
associated holonomy. This observation explains the fact that the
diffeomorphism constraint is equivalent to the Gau\ss\ constraint in
those cases. For $n^{(1)}=0$, on the other hand, there is no inner
automorphism in the transversal plane and no diffeomorphism constraint
is needed.

\section{Conclusion}

In this first part we presented the basic setting for a study of
cosmological models within loop quantum gravity. The kinematical level
was almost completely solved. It remains to determine the quantum
states which solve the Higgs constraint. This will be done in the next
part, together with a quantization of volume operators for
cosmological models.

In this early stage, of course, no physical statements concerning
features of a quantum theory of gravity in a cosmological context can
be made. Instead we concentrated on an application of these models as
test models for the general framework of quantum symmetry reduction
presented in reference \cite{SymmRed}. Cosmological models are well
suited for that task, because in the quantum formulation they consist
of a Higgs vertex only. This allows us to study these vertices, the
treatment of which has not yet been addressed in the general framework
(for a non-trivial Higgs constraint). In the next part we will
complete the solution of the Higgs constraint for isotropic models,
and show how spin network techniques can be used on the solution
space. Furthermore, the models discussed here again illustrate the
role of the reduced gauge group and of the relaxing of a partial gauge
fixing in the quantum theory.

Once we have the reduced models and their complete kinematical Hilbert
spaces at our disposal, we can use them as test models for poorly
understood issues of the full theory. The history of physics provides
many examples of how important the role of models with high symmetry
can be. For most theories the only known exact solutions are highly
symmetric, and such solutions can provide insights into conceptual
issues. Therefore, it should be helpful to use symmetric states to
investigate problems of loop quantum gravity. The outstanding task is,
of course, to understand the Hamiltonian constraint. The action of a
candidate \cite{QSDI,QSDII} for its quantization is already
complicated in a single vertex, and for its full action one has to
take into account the whole spin network it acts on by creating new
edges. For cosmological models, there is just a single vertex. No new
edges can be created; only the spins can be changed. Thus already the
simple geometry implies a simplification (an example for such a
simplification is the discussion of group averaging of the
diffeomorphism group for Bianchi IX models presented in the preceding
section). Moreover, the Wheeler--DeWitt operator contains the volume
operator, whose eigenvalues are not known explicitly for complicated
vertices.  For isotropic models there are only specific vertices, and
we will see in the next part that the complete spectrum of the volume
operator can be calculated. This should further facilitate an
investigation of the reduced Hamiltonian constraint.  Lastly, we
mention that the classical isotropic solutions are known explicitly
and very simple, so that a comparison with the classical theory will
be more easy to achieve. A quantization of Wheeler--DeWitt operators
for homogeneous models will be presented in the third part
\cite{cosmoIII}.

\section*{Acknowledgements}

The author is grateful to H.\ Kastrup for comments. He also thanks the
DFG-Graduierten-Kolleg ``Starke und elektroschwache Wechselwirkung bei
hohen Energien'' for a PhD fellowship.


\begin{thebibliography}{10}

\bibitem{DeWitt}
B.~S.\ DeWitt,
\newblock Quantum Theory of Gravity. I. The Canonical Theory,
\newblock {\em Phys.\ Rev.} 160 (1967) 1113--1148

\bibitem{Misner}
C.~W.\ Misner,
\newblock Quantum Cosmology. I,
\newblock {\em Phys.\ Rev.} 186 (1969) 1319--1327

\bibitem{Giulini}
D.\ Giulini,
\newblock What is the Geometry of Superspace,
\newblock {\em Phys.\ Rev.\ D} 51 (1995) 5630--5635

\bibitem{Polchinski}
J.\ Polchinski,
\newblock {\em String Theory}, volume I and II,
\newblock Cambridge University Press, 1998

\bibitem{Effective}
E.~S.\ Fradkin and A.~A.\ Tseytlin,
\newblock Quantum String Theory Effective Action,
\newblock {\em Nucl.\ Phys.\ B} 261 (1985) 1--27

\bibitem{Background}
C.~G.\ Callan, D.\ Friedan, E.~J.\ Martinec, and M.~J.\ Perry,
\newblock Strings in Background Fields,
\newblock {\em Nucl.\ Phys.\ B} 262 (1985) 593--609

\bibitem{StringCos}
J.~E.\ Lidsey, D.\ Wands, and E.~J.\ Copeland,
\newblock Superstring Cosmology,
\newblock hep-th/9909061, 1999

\bibitem{Veneziano}
G.\ Veneziano,
\newblock Scale Factor Duality for Classical and Quantum Strings,
\newblock {\em Phys.\ Lett.\ B} 265 (1991) 287--294

\bibitem{Rov:Loops}
C.\ Rovelli,
\newblock Loop Quantum Gravity,
\newblock gr-qc/9710008

\bibitem{AreaVol}
C.\ Rovelli and L.\ Smolin,
\newblock Discreteness of area and volume in quantum gravity,
\newblock {\em Nucl.\ Phys.\ B} 442 (1995) 593--619, [gr-qc/9411005],
\newblock Erratum: {\em Nucl.\ Phys.\ B} 456 (1995) 753

\bibitem{Loll:Vol}
R.\ Loll,
\newblock Volume Operator in Discretized Quantum Gravity,
\newblock {\em Phys.\ Rev.\ Lett.} 75 (1995) 3048--3051, [gr-qc/9506014]

\bibitem{Area}
A.\ Ashtekar and J.\ Lewandowski,
\newblock Quantum Theory of Geometry I: Area Operators,
\newblock {\em Class.\ Quantum Grav.} 14 (1997) A55--A82, [gr-qc/9602046]

\bibitem{Vol2}
A.\ Ashtekar and J.\ Lewandowski,
\newblock Quantum Theory of Geometry II: Volume Operators,
\newblock {\em Adv.\ Theo.\ Math.\ Phys.} 1 (1997) 388--429, [gr-qc/9711031]

\bibitem{Len}
T.\ Thiemann,
\newblock A length operator for canonical quantum gravity,
\newblock {\em J.\ Math.\ Phys.} 39 (1998) 3372--3392, [gr-qc/9606092]

\bibitem{Weaves}
A.\ Ashtekar, C.\ Rovelli, and L.\ Smolin,
\newblock Weaving a Classical Metric with Quantum Threads,
\newblock {\em Phys.\ Rev.\ Lett.} 69 (1992) 237--240

\bibitem{Smolin}
L.\ Smolin,
\newblock Time, Measurement and Information Loss in Quantum Cosmology,
\newblock gr-qc/9301016, 1993

\bibitem{AnoFree}
T.\ Thiemann,
\newblock Anomaly-free formulation of non-perturbative,
  four-dimensional Lorentzian quantum gravity,
\newblock {\em Physics Letters B} 380 (1996) 257--264, [gr-qc/9606088]

\bibitem{QSDI}
T.\ Thiemann,
\newblock Quantum Spin Dynamics {(QSD)},
\newblock {\em Class.\ Quantum Grav.} 15 (1998) 839--873, [gr-qc/9606089]

\bibitem{QSDII}
T.\ Thiemann,
\newblock Quantum Spin Dynamics {(QSD) II}: The Kernel of the
  Wheeler--DeWitt Constraint Operator,
\newblock {\em Class.\ Quantum Grav.} 15 (1998) 875--905, [gr-qc/9606090]

\bibitem{SymmRed} M.\ Bojowald and H.~A.\ Kastrup, 
\newblock Quantum
  Symmetry Reduction for Diffeomorphism Invariant Theories of
  Connections,
\newblock hep-th/9907042

\bibitem{cosmoIII}
M.\ Bojowald,
\newblock Loop Quantum Cosmology III: Wheeler--DeWitt Operators,
\newblock in preparation

\bibitem{KobNom}
S.\ Kobayashi and K.\ Nomizu,
\newblock {\em Foundations of Differential Geometry}, volume~1
(John Wiley \& Sons, New York 1963) chapter II.11;
volume~2 (New York 1969) chapter X

\bibitem{FermionHiggs}
T.\ Thiemann,
\newblock Kinematical Hilbert Spaces for Fermionic and Higgs Quantum Field
  Theories,
\newblock {\em Class.\ Quantum Grav.} 15 (1998) 1487, [gr-qc/9705021]

\bibitem{cosmoII}
M.\ Bojowald,
\newblock Loop Quantum Cosmology II: Volume Operators,
\newblock gr-gc/9910104

\bibitem{classAB}
M.~A.~H.\ MacCallum and A.~H.\ Taub,
\newblock Variational Principles and Spatially-Homogeneous Universes,
Including Rotation, 
\newblock {\em Commun.\ math.\ Phys} 25 (1972) 173--189

\bibitem{midisup}
C.~G.\ Torre,
\newblock Midi-Superspace Models of Canonical Quantum Gravity,
\newblock {\em Int.\ J.\ Theor.\ Phys.} 38 (1999) 1081--1102, 
  [gr-qc/9806122]

\bibitem{MacCallum}
G.~F.~R.\ Ellis and M.~A.~H.\ MacCallum,
\newblock A Class of Homogeneous Cosmological Models,
\newblock {\em Commun.\ Math.\ Phys.} 12 (1969) 108--141

\bibitem{MiniQuant}
A.\ Ashtekar, R.\ Tate, and C.\ Uggla,
\newblock Minisuperspaces: Observables and Quantization,
\newblock {\em Int.\ J.\ Mod.\ Phys.} D2 (1993) 15--50, [gr-qc/9302027]

\bibitem{Kodama}
H.\ Kodama,
\newblock Specialization of Ashtekar's Formalism to Bianchi Cosmology,
\newblock {\em Prog.\ Theo.\ Phys.} 80 (1988) 1024--1040

\bibitem{AshVarReell}
J.~F.\ Barbero~G.,
\newblock Real Ashtekar variables for Lorentzian signature space-times,
\newblock {\em Phys.\ Rev.\ D} 51 (1995) 5507--5510, [gr-qc/9410014]

\bibitem{Ellis}
G.~F.~R.\ Ellis,
\newblock Dynamics of Pressure-Free Matter in General Relativity,
\newblock {\em J.\ Math.\ Phys.} 8 (1967) 1171--1194

\bibitem{Kubyshin}
Yu.~A.\ Kubyshin, O.\ Richter, and G.\ Rudolph,
\newblock Invariant Connections on Homogeneous Spaces,
\newblock {\em J.\ Math.\ Phys} 34 (1993) 5268--5282

\bibitem{ALMMT} A.\ Ashtekar, J.\ Lewandowski, D.\ Marolf, J.\ 
  Mour\~ao, and T.\ Thiemann,
\newblock Quantization of diffeomorphism
  invariant theories of connections with local degrees of freedom,
\newblock {\em J.\ Math.\ Phys.} 36 (1995) 6456--6493,
  [gr-qc/9504018]

\bibitem{QSDIII} T.\ Thiemann, 
\newblock {QSD III}: Quantum Constraint
  Algebra and Physical Scalar Product in Quantum General Relativity,
\newblock {\em Class.\ Quantum Grav.} 15 (1998) 1207--1247,
  [gr-qc/9705017]

\end{thebibliography}
\end{document}